\documentclass[journal]{vgtc}                     

\onlineid{1142}

\vgtccategory{Analytics \& Decisions}

\vgtcpapertype{System}

\title{Compress and Compare: Interactively Evaluating Efficiency and Behavior Across ML Model Compression Experiments}

\author{%
  \authororcid{Angie Boggust*\textsuperscript{\dag}}{1234-5678-9012},
  \authororcid{Venkatesh Sivaraman*\textsuperscript{\dag}}{1234-5678-9012},
  \authororcid{Yannick Assogba}{1234-5678-9012},
  \authororcid{Donghao Ren}{1234-5678-9012},
  \\
  \authororcid{Dominik Moritz}{1234-5678-9012}, and
  \authororcid{Fred Hohman}{0000-0002-4164-844X}
}

\authorfooter{
  \item \textup{*} Authors contributed equally.
  \item \textup{\dag} Work done at Apple.
  \item
  	Angie Boggust is with the Massachusetts Institute of Technology.
  	E-mail: aboggust@csail.mit.edu
  \item
  	Venkatesh Sivaraman is with Carnegie Mellon University. E-mail: venkats@cmu.edu

  \item Yannick Assogba, Donghao Ren, Dominik Moritz, and Fred Hohman are with Apple. E-mail: \{yassogba,donghao,domoritz,fredhohman\}@apple.com
}

\abstract{%
To deploy machine learning models on-device, practitioners use compression algorithms to shrink and speed up models while maintaining their high-quality output.
A critical aspect of compression in practice is model comparison, including tracking many compression experiments, identifying subtle changes in model behavior, and negotiating complex accuracy-efficiency trade-offs. 
However, existing compression tools poorly support comparison, leading to tedious and, sometimes, incomplete analyses spread across disjoint tools.
To support real-world comparative workflows, we develop an interactive visual system called \system. 
Within a single interface, \system surfaces promising compression strategies by visualizing provenance relationships between compressed models and reveals compression-induced behavior changes by comparing models' predictions, weights, and activations.
We demonstrate how \system supports common compression analysis tasks through two case studies, debugging failed compression on generative language models and identifying compression artifacts in image classification models. 
We further evaluate \system in a user study with eight compression experts, illustrating its potential to provide structure to compression workflows, help practitioners build intuition about compression, and encourage thorough analysis of compression’s effect on model behavior.
Through these evaluations, we identify compression-specific challenges that future visual analytics tools should consider and \system visualizations that may generalize to broader model comparison tasks.

}


\teaser{
  \centering
  \includegraphics[width=\linewidth]{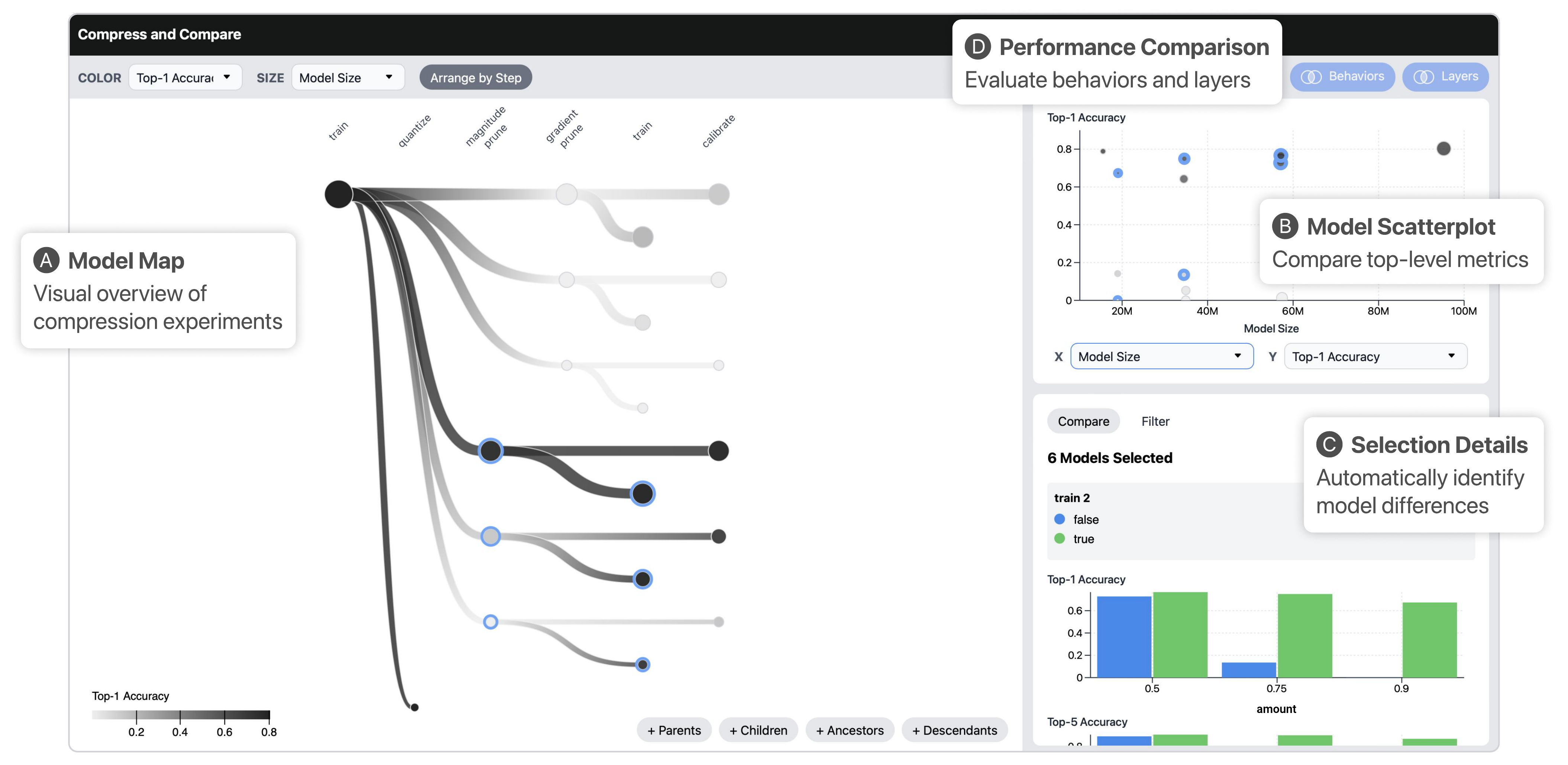}
  \caption{
  \system helps ML practitioners analyze and compare compression experiments. The (A) \modelmap displays an overview of the compressed models and the experimental operations used to create them. Users can compare models' top-level metrics in the (B) \scatterplot and their operational differences in the (C) \selection view. To compare models' behaviors and internal layer characteristics, users can visit the (D) \performancecomparison views shown in \cref{fig:behaviors-layers-view}.
  }
  \label{fig:teaser}
}

\graphicspath{{figs/}{figures/}{pictures/}{images/}{./}} 

\usepackage{tabu}                      
\usepackage{booktabs}                  
\usepackage{lipsum}                    
\usepackage{mwe}                       

\usepackage{mathptmx}                  

\usepackage{enumitem}

\usepackage{arydshln} 
\usepackage{xspace,xpunctuate}

\usepackage{amsmath}

\makeatletter
\newcommand{\dummylabel}[2]{\def\@currentlabel{#2}\label{#1}}
\makeatother

\begin{document}

\dummylabel{app:squad-models}{S2}
\dummylabel{app:celeba-models}{S3}
\dummylabel{fig:userstudy-model-details}{S1}
\dummylabel{tab:model-training}{S1}
\dummylabel{tab:imagenet-models}{S2}
\dummylabel{fig:squad-model-details}{S2}
\dummylabel{fig:squad_layers}{S3}


\definecolor{RoyalBlue}{HTML}{0071BC}
\newcommand{\todo}[1]{{\textcolor{RoyalBlue}{[#1]}\normalfont}}
\newcommand{\angie}[1]{{\textcolor{RoyalBlue}{[#1 -AB]}\normalfont}}
\newcommand{\venkat}[1]{{\textcolor{RoyalBlue}{[#1 -VS]}\normalfont}}
\newcommand{\yannick}[1]{{\textcolor{RoyalBlue}{[#1 -YA]}\normalfont}}
\newcommand{\donghao}[1]{{\textcolor{RoyalBlue}{[#1 -DR]}\normalfont}}
\newcommand{\dom}[1]{{\textcolor{RoyalBlue}{[#1 -DM]}\normalfont}}
\newcommand{\fred}[1]{{\textcolor{RoyalBlue}{[#1 -FH]}\normalfont}}

\definecolor{SystemPurple}{HTML}{00BDB4}
\definecolor{SystemBlue}{HTML}{007aff}
\definecolor{SystemRed}{HTML}{ff3b30}
\newcommand{\add}[1]{{\textcolor{SystemBlue}{#1}\normalfont}}
\newcommand{\remove}[1]{{\textcolor{SystemRed}{\st{#1}}\normalfont}}

\newcommand{\ie}{{i.e.,}\xspace}
\newcommand{\eg}{{e.g.,}\xspace}
\newcommand{\ea}{{et~al\xperiod}\xspace}
\newcommand{\aka}{{a.k.a.}\xspace}
\newcommand{\etc}{{etc\xperiod}\xspace}
\newcommand{\etal}{{et al\xperiod}\xspace}

\definecolor{urlBlue}{HTML}{007aff}
\newcommand{\urlColor}[1]{{\color{urlBlue}{#1}\normalfont}}
\newcommand{\codeURL}{\url{ https://github.com/apple/ml-compress-and-compare}\xspace}

\newcommand{\location}{{a large consumer technology company}\xspace}
\newcommand{\system}{\textsc{Compress and Compare}\xspace}
\newcommand{\NumStudyParticipants}{8\xspace}
\newcommand{\NumStudyParticipantsWord}{eight\xspace}
\newcommand{\behaviors}{\textbf{Behaviors}\xspace}
\newcommand{\layers}{\textbf{Layers}\xspace}
\newcommand{\overview}{\textbf{Compression Overview}\xspace}
\newcommand{\scatterplot}{\textbf{Model Scatterplot}\xspace}
\newcommand{\filter}{\textbf{Filter}\xspace}
\newcommand{\modelmap}{\textbf{Model Map}\xspace}
\newcommand{\performancecomparison}{\textbf{Performance Comparison}\xspace}
\newcommand{\selection}{\textbf{Selection Details}\xspace}


\crefname{table}{Table}{Tables}
\Crefname{table}{Table}{Tables}


\firstsection{Introduction}

\maketitle

\label{sec:introduction}
Machine learning (ML) models have dramatically increased in scale over the past several years, with published models rising from 1 billion parameters in 2018 to over 100 billion parameters as of 2024~\cite{villalobos2022machine, owid2022artificialintelligence}.
This trend has produced models with exciting emergent capabilities that have enabled new user experiences, like real-time translation~\cite{openai2024gpt4o} and code generation~\cite{friedman2022introducing}. 
However, this scale also incurs greater technical, financial, and environmental costs to integrate these models into everyday use~\cite{bender2021on}. 
As a result, \textit{model compression} has emerged as an essential family of techniques to make large models viable for practical usecases, particularly in domains where models must run on end-user devices to lower latency or access private user data~\cite{hohman2024compression}.

ML practitioners apply compression with the intent to maintain the accuracy of a large model while reducing the space required to store it and the time required to perform inference. 
However, which compression technique or combination of techniques will achieve this balance remains task- and model-specific~\cite{hohman2024compression}. 
The ML literature has proposed various compression techniques for different model architectures and user priorities, such as low space consumption, low latency, or fast execution on optimized hardware~\cite{gholami2022survey,yu2022unified,frantar2023massive,frankle2018lottery}. 
Nevertheless, identifying the right compression strategy can require anywhere from a few to several dozen experiments~\cite{hohman2024compression}, taking time that is often not accounted for in accuracy-focused model development timelines. 
It can also be challenging to communicate experimental results within and across teams, particularly those with varying ML expertise. 
Even when these efforts are successful, compression can alter model behavior in subtle and unexpected ways, creating new errors or biased outputs~\cite{hooker2020characterising} that are hard to capture with a single metric.

Although compression is increasingly used in research and industry domains, there has been little work using visualization to make compression techniques more interpretable and comprehensible. 
Initial work on compression visualization has focused on specific compression techniques, like neural network pruning~\cite{li2020cnnpruner,schlegel2022vinnpruner} or profiling a single model's power and performance characteristics on specific hardware~\cite{hohman2024talaria}. 
While these methods begin to demonstrate the value of visual tools for compression tasks, ML practitioners often need to take a broader approach to experimentation.
As they mix and match multiple techniques in different orderings, the number of experiments and models they produce quickly expands, creating visualization challenges that are not well-supported by available tools for either interactive compression or model comparison~\cite{gleicher2020boxer,murugesan2019deepcompare,narkar2021model}.

In this work, we explore how to address model compression challenges using interactive visualization. 
We first identify four model compression challenges by synthesizing prior qualitative findings on how ML practitioners use compression~\cite{hohman2024compression} with insights from the ML literature.
In response to these challenges, we introduce \system, an interactive visualization system for comparing the performance and behavior of a suite of compressed models. 
Through an overview visualization called the \modelmap, our system helps users track their compression experiments and how they relate to one another.
Users can select subsets of models to automatically visualize differences in their accuracy, efficiency, and provenance. 
To deeply inspect a smaller set of models, the system provides detailed comparisons of instance-level behaviors and internal activations. 
Through case studies and user studies, we demonstrate how \system can lead to insights throughout model compression and how it expands the design space of interactive ML development tools to account for challenges made salient by compression. 
We contribute:

\begin{itemize}
\item \textbf{Four identified compression challenges} ML practitioners face when developing and selecting model compression strategies.
\item \textbf{\system}, an interactive visualization system enabling comparative analysis over many compressed models.
\item \textbf{Case studies on two common compression tasks} demonstrating how \system can help debug failed compression experiments and identify compression-induced bias. 
\item \textbf{A user study with \NumStudyParticipantsWord compression practitioners} illustrating how \system help users build intuition by providing structure to their compression workflows.
\end{itemize}

\section{Background and Related Work}
\label{sec:related-work}
\subsection{Techniques for Model Compression}

Model compression encompasses various techniques that reduce the storage space, memory, power, or time required to run an ML model while preserving its original behavior as much as possible~\cite{menghani2023efficient, choudhary2020comprehensive, yu2018model, deng2020model, treviso2023efficient}. 
Compression is increasingly essential for running ML models, both in resource-constrained settings such as mobile devices and for extremely large models (e.g., generative language models).

Most compression techniques fall into one of three classes: (1) \textit{quantization and palettization} reduce the space required to store each individual parameter, (2) \textit{pruning} removes parameters while optionally adjusting the others to compensate, and (3) \textit{factorization and distillation} find a different set of parameters that mimic the behavior of the original model~\cite{choudhary2020comprehensive}. The most straightforward instantiations of these techniques are quantization (converting high-precision formats like 32-bit floats to lower-precision formats like 8-bit integers) and unstructured magnitude pruning (zero-ing out weights with the smallest absolute values). These foundational techniques form the starting point for many real-world compression strategies because they perform well in practice, are easy to understand, and are straightforward to compute~\cite{liu2018rethinking,hohman2024compression}. Additional routines often employed to tune the resulting compressed model include \textit{fine-tuning} (training the model on a data subset) and \textit{calibration} (adjusting model parameters to compensate for compression).

In cases where off-the-shelf techniques are insufficient, task-specific techniques have been developed to achieve better efficiency trade-offs~\cite{gamboa2020campfire,yu2018nisp,lee2018snip,gale2019state,he2018soft,zhu1710prune}. These methods vary by which aspects of model efficiency they target, how computationally expensive applying the compression is, and whether or not they depend on additional training or calibration data. For example, some methods utilize random data samples to decide which parameters can most easily be pruned or restore intermediate activations~\cite{baykal2019sipping,molchanov2019importance,frantar2023massive}, while others are data-agnostic~\cite{han2015deep,renda2020comparing} and can optionally be followed by a retraining step. Individual weights can be modified independently~\cite{renda2020comparing,frantar2023massive}, or compression can be performed in a structured manner at the level of neurons or layers~\cite{suau2020filter}. These algorithmic choices give rise to a large space of possible compression strategies, each of which has different overall performance characteristics in terms of space consumption, inference time, and accuracy preservation. Helping ML practitioners navigate this space is a key design opportunity addressed in our work (see \cref{sec:formative}).


\subsection{Pitfalls in Evaluating Compressed Models}
While compression techniques are designed to improve model efficiency while preserving accuracy, they have been known to substantially alter model behavior even while maintaining similar top-level metrics.
For example, Hooker~\etal~\cite{hooker2020characterising} find that pruning image classification models has negligible effects on overall accuracy but disproportionately impacts the accuracy of rare subgroups.
Similarly, Liebenwein~\etal~\cite{liebenwein2021lost} find that pruning image models often results in poor generalization on distribution-shifted inputs.

Since these behavior changes do not always impact top-level metrics, they can be hard to identify in advance without conducting bespoke analyses dedicated to finding them. 
For example, a prior interview study with ML practitioners~\cite{hohman2024compression} described a situation where an object detection model produced jittered outputs after quantization, a phenomenon that was not uncovered until the model was tested on-device in a demo setting. 
While tools for model comparison are applicable to this task, compression poses additional analytical challenges, such as comparing more than two models at once and understanding how models are derived from one another. Our work aims to address these challenges by providing practitioners with visual tools to inspect the behavior of several compressed models during development.

\subsection{Visualization for Model Understanding and Comparison}

Visualization has been essential in building generalizable knowledge about ML architectures~\cite{hohman2018visual,wang_visual_2024, yuan2021survey} and helping practitioners make sense of specific models~\cite{kahng2018activis,bauerle2022symphony,DNIKit,robertson2023angler,wu2019errudite,suresh2023kaleidoscope}. 
Many tools use comparison to make insights about model behavior and internals more meaningful, \eg by jointly visualizing embedding spaces to distinguish meaningful data clusters from spurious ones~\cite{sivaraman2022emblaze,boggust2022embedding}. 
Other comparative approaches extend analysis subtasks to multiple models, including comparing model errors~\cite{narkar2021model}, instance-level outputs~\cite{gleicher2020boxer,xuan2022vaccnn}, or internal representations~\cite{murugesan2019deepcompare}. 
Tools have been developed to help practitioners select from a wide array of models using comparisons of top-level metrics~\cite{Schelter2017,zaharia2018accelerating}. 
In our work, we use comparative visualization to help users identify promising strategies and filter out unviable experiments.


Generating and evaluating \textit{compressed} ML models is a more nascent area in VIS4ML\,---\,most prior work for this task is limited to either specific compression techniques or profiling efficiency metrics without considering behavior. For example, \textit{CNN}Pruner~\cite{li2020cnnpruner} uses a Taylor expansion criterion to prune filters in convolutional networks, while ViNNPruner~\cite{schlegel2022vinnpruner} supports a wider variety of architectures but uses an interactive pruning scheme that is difficult to scale to large models. Meanwhile, Talaria~\cite{hohman2024talaria} helps users estimate the effects of compression on the efficiency of any model, but it requires accuracy to be evaluated separately. Unlike these prior works, our system aims to support more general, iterative compression workflows that may involve dozens of models with different sets of techniques applied.

Overall, these prior systems have focused on either model compression or model comparison alone without taking into account how the two tasks are often intertwined during a model development pipeline~\cite{hohman2024compression}. Evaluating the potential of interactive tools to help at the intersection of these two challenges is the primary focus of our work.

\section{Design Challenges for Compression}
\label{sec:formative}

To identify key compression challenges and motivate the design of interactive tools for compression, we synthesized insights from ML and HCI literature.
Recently, Hohman~\etal~\cite{hohman2024compression} explored ML practitioners' needs and perspectives on model compression via an interview study with 30 compression experts.
This work describes how compression experts experiment with different compression techniques to satisfy efficiency and accuracy constraints within multidisciplinary teams.
While their focus was providing a broad overview of compression experts' tacit knowledge, we distill specific findings from their study that are relevant to the design of compression tools and supplement them with recent insights from ML literature to synthesize key challenges for our system to address.

\begin{figure}
    \centering
    \includegraphics[width=0.95\linewidth]{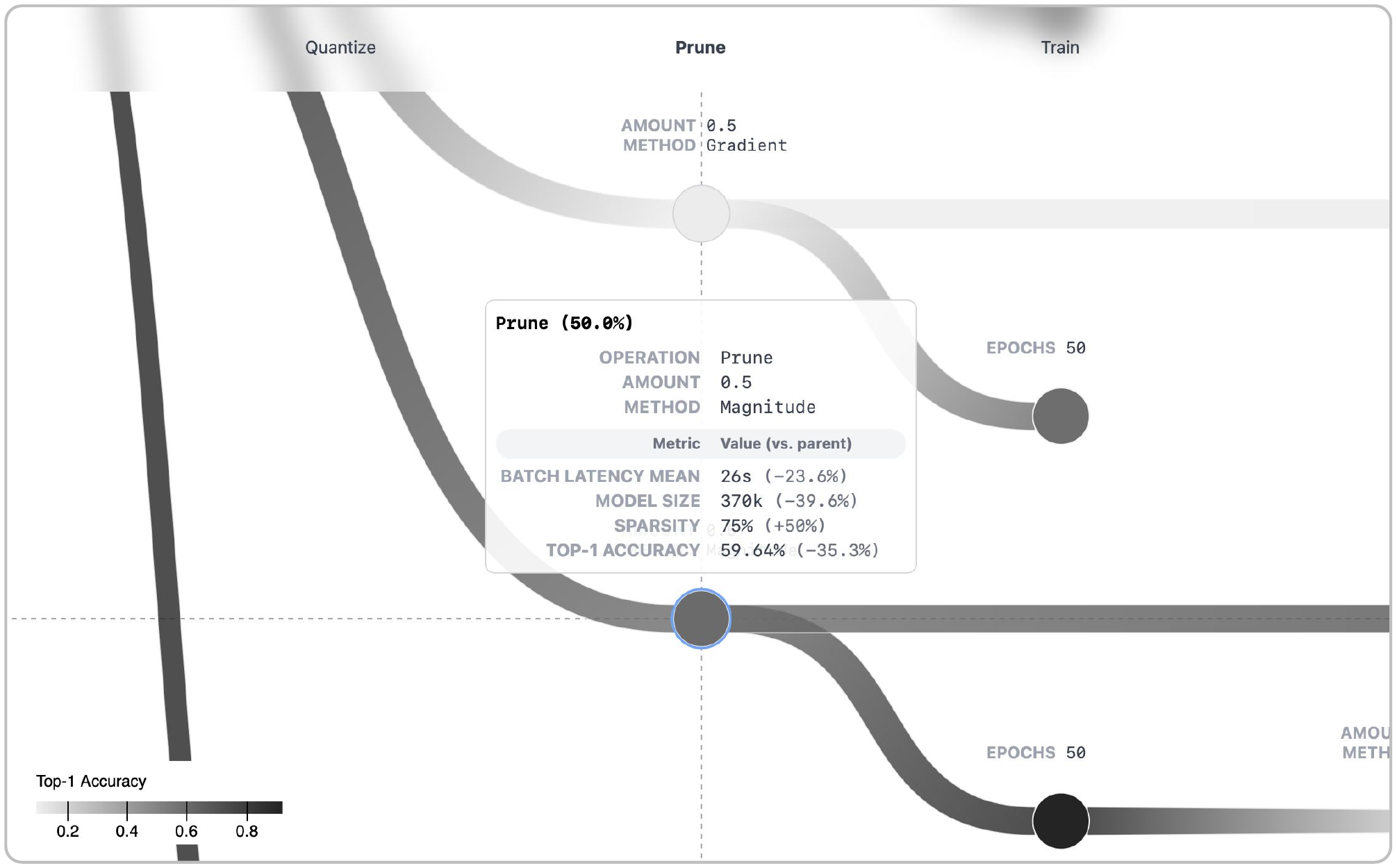}
    \caption{Hovering over a \modelmap model displays a tooltip containing the models' top-level metrics, including latency, size, sparsity, accuracy, and compression operation. Here, the selected model has been $50\%$ pruned, improving its latency and size but reducing its accuracy.}
    \label{fig:tooltip}
\end{figure}

\begin{enumerate}[label={\bfseries C\arabic*.}, ref={\bfseries C\arabic*},itemsep=1ex, labelindent=0pt, wide=0pt]

\item \textit{Identifying the optimal compression strategy is time-consuming and task-specific.}  
A one-size-fits-all strategy for model compression does not exist.
Even compression experts do not know how to achieve the best balance of efficiency and accuracy a priori and often experiment with multiple compression algorithms for each new task and model~\cite{hohman2024compression, hohman2024talaria}.
However, existing compression tools focus on exploring the results of a single compression experiment~\cite{weidele2020autoaiviz,ono2020pipelineprofiler,Schelter2017} instead of assessing the design space of all possible experiments.
As a result, compression experts in our evaluation study (\cref{sec:evaluation}) had complex and time-consuming comparison workflows, such as flipping between the same tool loaded with different models and maintaining large spreadsheets of model results.
Compression tools should support practitioners in comparing compressed models based on their performance, efficiency, behavior, and provenance to identify promising strategies.
\label{challenge:no-single-method}
\item \textit{Compression requires human trade-offs across multiple metrics.}
Practitioners often discuss model compression as an effort to meet resource targets on memory, time, and accuracy~\cite{hohman2024compression}. 
These budgets are often negotiated, set, and adjusted based on how models would affect the user experience. 
For example, a model that could run overnight while a device is charging would be granted more memory and time since it would be unlikely to disrupt the user. 
Similar decisions take place around accuracy; a model that errs more on sensitive or critical subgroups would not be deployed, while one that makes reasonable or recoverable errors could be viable~\cite{hohman2024compression}.
As a result, practitioners need tools to quickly assess model variants and make trade-offs between their metrics to help them satisfy these multifaceted constraints.
\label{challenge:budgeting}

\begin{figure}
    \centering
    \includegraphics[width=\linewidth]{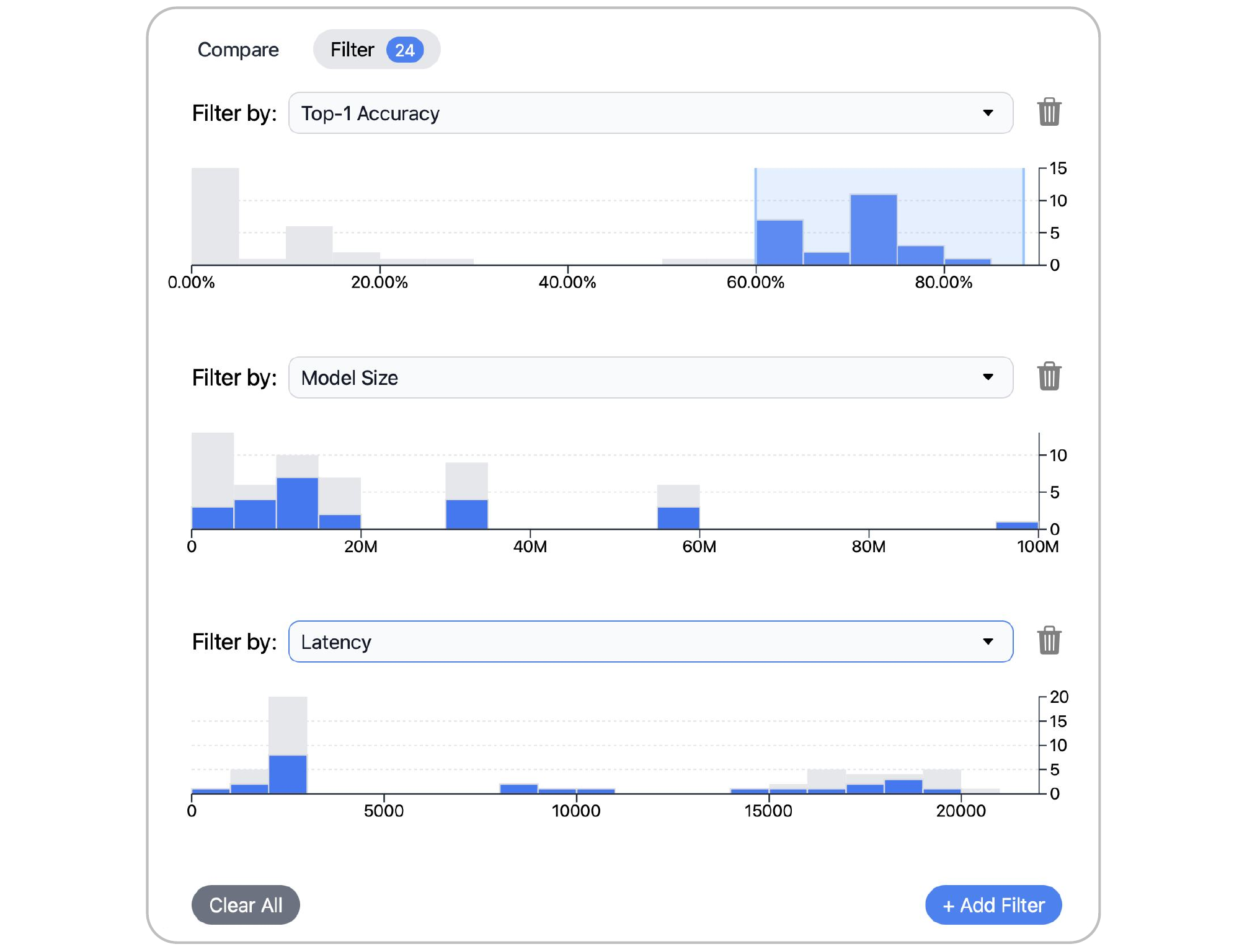}
    \caption{The \filter view allows users to define hard budgets using metrics, like accuracy, model size, or latency. The user can add several filters, see the distribution of values for the selected metrics, and brush on the histograms to disable models with metric values outside of the range.}
    \label{fig:filter-view}
\end{figure}
\begin{figure*}[t]
    \centering
    \includegraphics[width=\linewidth]{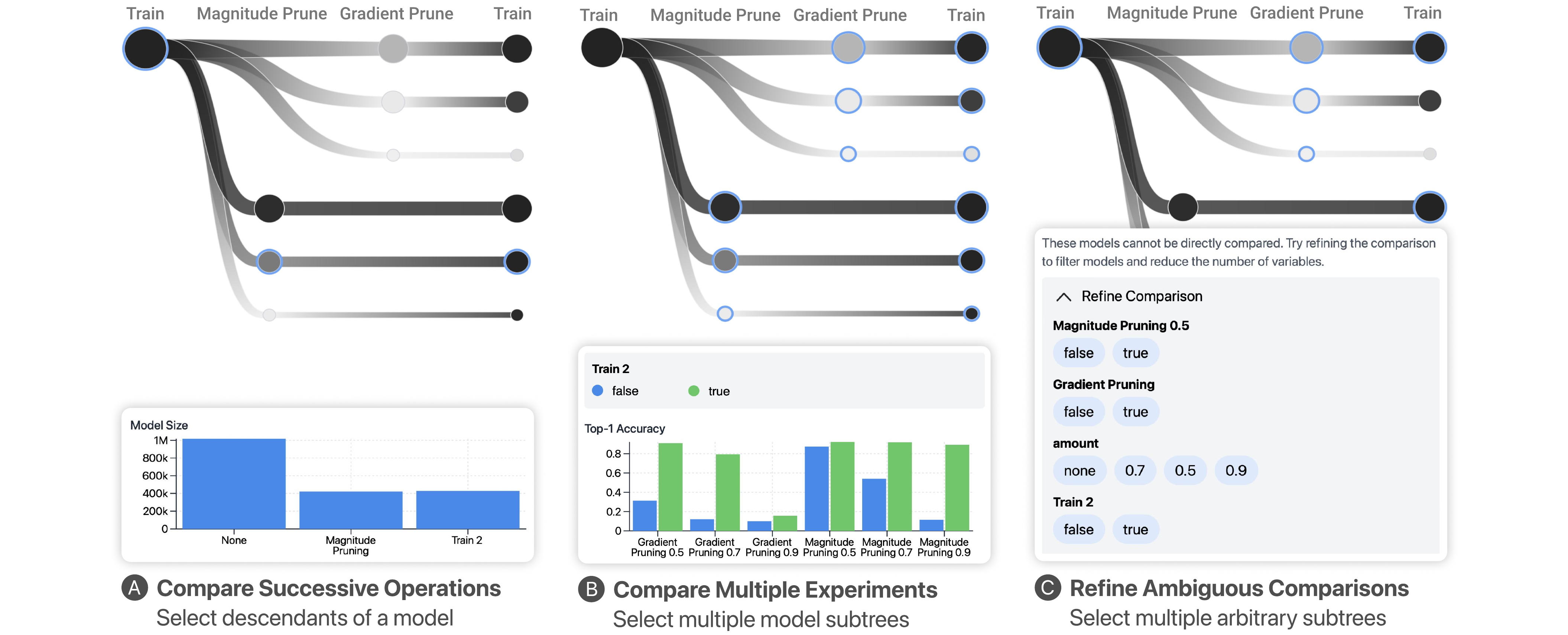}
    \caption{
    The \selection view automatically generates meaningful comparisons between selected \modelmap models, such as comparing successive operations by selecting the descendants of a model (A) or comparing multiple compression algorithms by selecting multiple subtrees (B). If a direct comparison does not exist, the interface prompts the user to refine the comparison to the attributes they are interested in visualizing (C).
    }
    \label{fig:selection-examples}
\end{figure*}

\item \textit{Top-level metrics can obscure important differences between compressed models.}
Although compression can achieve comparable performance to the original model, it can alter the model's behavior by introducing new errors or biases.
In some cases, these differences are random or imperceptible, but in others, they can be problematic, such as reducing the quality of model predictions~\cite{hohman2024compression}, changing model explanations~\cite{liebenwein2021lost}, or increasing bias~\cite{hooker2020characterising}.
Unfortunately, behavioral changes do not always correspond to a change in evaluation metrics, so humans must be involved in the evaluation process to catch dangerous behaviors. 
While various visualization techniques compare the behaviors of \textit{pairs} of models~\cite{murugesan2019deepcompare,boggust2022embedding,sivaraman2022emblaze}, there remain opportunities to help practitioners compare \textit{many} compressed models' behavioral changes.
\label{challenge:behavior-differences}


\item \textit{Compression can have unintended, hard-to-debug effects on model internals.}
Practitioners often have a set of heuristics they expect compression algorithms to follow, such as not compressing early layers and compressing layers proportional to their number of parameters.
However, when compression algorithms do not follow these expectations, it can be difficult for practitioners to determine why~\cite{hohman2024compression}.
For example, a network's outputs may change significantly because a layer ceased to produce meaningful output or because its activations had a different distribution that caused downstream layers' outputs to change as well.
To ensure compression only impacts the desired portions of the model, practitioners often go layer-by-layer to find errors and bottlenecks. 
Particularly for deep networks with hundreds of layers and billions of intermediate outputs, parameter-wise model comparison becomes a challenging task and an opportunity for visualization tooling.
\label{challenge:internal-differences}



\end{enumerate}

\section{Design of \system}
\label{sec:system}

We developed an interactive interface called \system to address the four compression design challenges (\cref{sec:formative}).
The tool consists of two main views: the \overview supports high-level comparison and model selection from large-scale compression experiments (\ref{challenge:no-single-method} and \ref{challenge:budgeting}), and the \performancecomparison view enables fine-grained inspection of model behaviors and internals for a small number of candidate models (\ref{challenge:behavior-differences} and \ref{challenge:internal-differences}). 

\subsection{Compression Overview}
\label{sec:model-map}

In \system, experiments are represented as a set of trees, where each node is a model and edge is an \textit{operation} performed on a parent model to produce a child model. 
This structural choice helps address one of the main visualization challenges of tracking compression experiments (\ref{challenge:no-single-method}): simultaneously depicting the variation in metrics across models along with dependencies in how the models were generated. 
The \modelmap (\cref{fig:teaser}A) addresses this using a node-link tree diagram, where nodes are positioned using a custom algorithm that vertically aligns nodes based on either the operations used to produce them or their step in the compression experiment.
Models are rendered as circles whose color and size encode performance properties, commonly accuracy and model size. 
To emphasize the sequential nature of compression experimentation, the color and width of the edges smoothly interpolate between the parent and child nodes.
Hovering over a model displays a tooltip with the model's top-level metrics (\cref{fig:tooltip}), such as latency, size, sparsity, accuracy, and the operation that created this new model from its parent.

While the \modelmap's layout prioritizes understanding model dependencies, the \scatterplot (\cref{fig:teaser}B) visualizes model metrics along the spatial axes, helping direct the user to viable models and find natural model groupings. 
For example, the classic Pareto curve often used by compression experts~\cite{hohman2024compression} can easily be recreated by setting model size or latency on the $x$-axis and accuracy on the $y$-axis. 
Node color and size are consistent between the \scatterplot and \modelmap, and the two visualizations are connected via brushing and linking~\cite{becker_brushing_1987}.
The \filter view also depicts model metrics through customizable histograms that can be brushed to filter the \modelmap and \scatterplot.
When the user specifies a filter, models that do not meet the filter criteria become semitransparent and unselectable. 
This allows users to narrow down the set of viable models by expressing their project's space and performance budgets (\ref{challenge:budgeting}).

When one or more models are selected, the user can view information about the models' metrics in the \selection view (\cref{fig:teaser}C). 
While displaying metrics for multiple models in a table would be straightforward, it would obscure the dependency structure between the models, making it harder to reason about the effects of different compression operations. 
Therefore, we develop a technique to automatically create a grouped bar chart from a subset of the model dependency tree. 
Our algorithm traverses the tree recursively to identify a minimum-cost set of ``variables'' that compactly explain the selected models' differences. 
Variables include operation parameter values, presence or absence of an operation, and type of operation applied. 
The algorithm attempts to fit multiple alternative variable types at each stage of the tree traversal and chooses the assignment that results in the shortest and simplest set of variables (\eg a variable for an operation's parameters is considered simpler than a variable for operation type). 
This ensures that similar operations are mapped to each other.
For example, the models resulting from \textit{$\text{Prune} \rightarrow \text{Quantize}$} and \textit{$\text{Prune} \rightarrow \text{Calibrate} \rightarrow \text{Quantize}$} can be explained by \textit{Calibrate = true or false}.

If the models can be represented using two variables or fewer, we generate a bar chart by mapping the $x$-axis and color encodings to the two variables. 
If more variables are required, the algorithm attempts to iteratively simplify the variable set by identifying conditional dependencies and cumulative relationships between pairs of variables.
As shown in \cref{fig:selection-examples}A and B, this simplification allows us to visualize several successive operations as a single $x$-axis encoding or combine multiple variables that are related to the same operation. 
When the list of variables needed to describe the selection cannot be simplified to two encodings, a Refine Comparison view allows the user to generate bar charts for subsets of the selection that \textit{can} be compared (\cref{fig:selection-examples}C).

\subsection{Performance Comparison}
\label{sec:performance-comparison}

\begin{figure*}
    \centering
    \includegraphics[width=\linewidth]{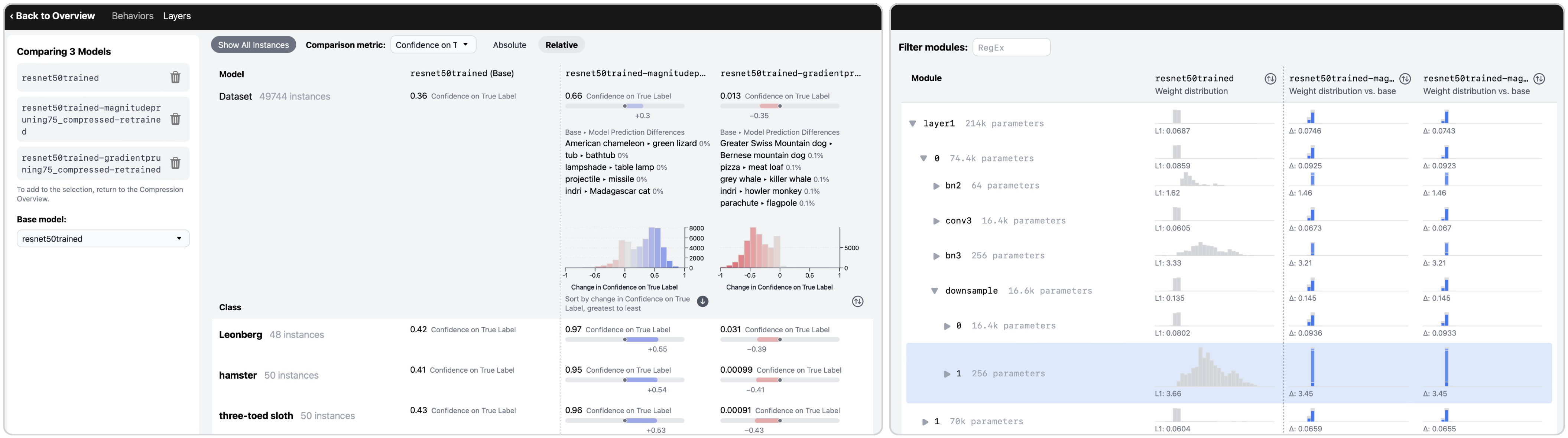}
    \caption{
    The \performancecomparison view provides an in-depth comparison of two or more models.
    The \behaviors tab (left) displays differences between models' predictions, distributions of comparison metrics, and a breakdown of the selected comparison metric at the class or instance level.
    Meanwhile, the \layers tab (right) compares the sparsity, weights, and activations across layers in the models using a file tree structure.
    }
    \label{fig:behaviors-layers-view}
\end{figure*}

While the \overview visualizes trends across large sets of models, the \performancecomparison view enables a deeper comparison of a smaller group of models' \behaviors and \layers.
We use a combination of juxtaposition and explicit encodings~\cite{gleicher2017considerations} to enable comparisons of multiple models at a time. 
Comparison is performed with respect to a base model, which is user-customizable but defaults to the selected model closest to the tree's root node.

The \behaviors tab (\cref{fig:behaviors-layers-view}, left) presents the results of evaluating each selected model on a validation dataset. 
Each model is a column in a table, where rows represent either class-level comparisons or individual instances. 
When configuring their data, users can define \textit{comparison metrics} that operate on the model outputs, enabling comparisons like differences in top-1 predictions or the KL divergence of the softmax probabilities. 
These comparison metrics are summarized in the table headers and are depicted as sparkline bar charts in each row. 
Notably, the interface supports both absolute per-model values and relative values compared to the base model. 
Users can sort and filter by absolute or relative metrics for any model, allowing them to quickly identify classes or instances impacted the most by compression (\ref{challenge:behavior-differences}). 

The final and lowest-level component of the interface is the \layers tab (\cref{fig:behaviors-layers-view}, right), which exposes the internals of the selected models.
Like the \behaviors tab, this view comprises a table where each column contains information about a model relative to the base; however, here, each row represents a module in the nested hierarchy of modules that makes up each network. 
Within this structure, the user can choose from visualizing the proportion of zero weights in each module, the distribution of the weight values, and the distribution of the activations (intermediate outputs) on a random data sample. 
Weight values and activations are depicted as stacked histograms so that the height of the bars forms the overall value distribution while the color indicates the degree of change relative to the base. 
This highlights parameters and models that have changed more than others, which can reveal bugs such as over-pruned layers or outlier activations (\ref{challenge:internal-differences}).

\subsection{Setup and Implementation Details}
\label{sec:data-ingestion}

We designed \system for a highly customizable user workflow. 
To begin visualizing models, users write a simple Python script that invokes the \system backend server and provides information about the models.
Users specify models as a JSON object that details the operations used to produce each model and their performance across a set of user-defined metrics.
Users can easily integrate specification creation into their existing model training procedures by updating the JSON file each time they train and evaluate a new model or evaluate against a new metric.
To access information about the model's behaviors and layers, users write Python callbacks to retrieve instance-level outputs and layer activations, both of which can be either pre-computed or evaluated in real-time.
This flexibility allows \system to support any Python-based ML toolchain and accommodate very large models.
Additionally, the \system Python package provides helper functions to accelerate setup with common frameworks such as PyTorch and HuggingFace.
The model servers for the use cases in this paper require around 200 lines of code, mostly consisting of boilerplate code that would already have been written in the course of experimentation.

The \system frontend, implemented in SvelteKit\footnote{\url{https://kit.svelte.dev}}, is static and can be hosted publicly.
Visualizations are developed using D3.js\footnote{\url{https://d3js.org}} and LayerCake\footnote{\url{https://layercake.graphics}}. 
Code is available at: \codeURL.

\section{Case Studies of Common Compression Tasks}
We illustrate how \system supports real-world compression workflows via two case studies.

\subsection{Repairing Models Broken By Compression}
\label{sec:case-study-broken-compression}

\begin{figure*}[t]
    \centering
    \includegraphics[width=\linewidth]{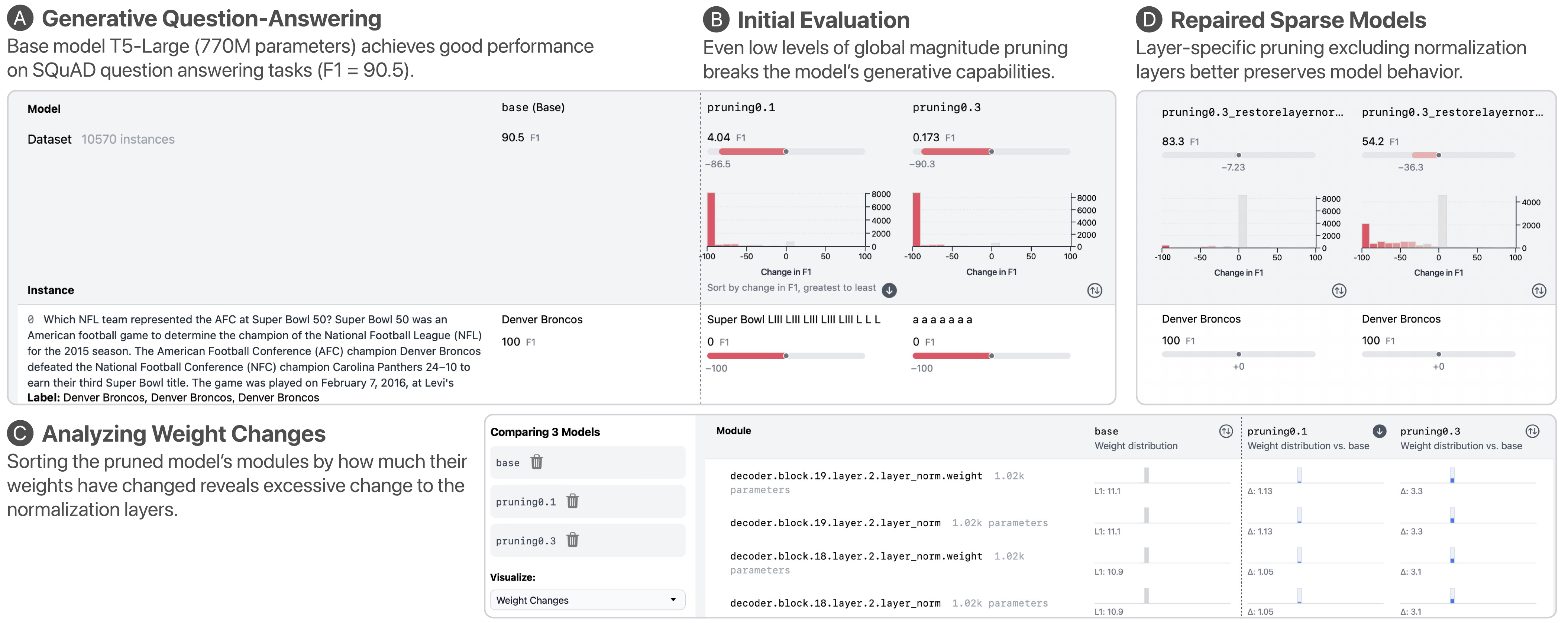}
    \caption{
    \system helps debug compression experiments. On a generative question-answering task (A), the \behaviors view reveals that global magnitude pruning severely deteriorates generation quality (B), whereas layer-specific pruning matches the original model's behavior (C).
    }
    \label{fig:squad-case-study}
\end{figure*}

A previously accurate model can ``break'' when compression is applied too heavily or broadly, resulting in low performance and nonsensical outputs.
However, it can be challenging for users to determine which components of a model are causing its performance to degrade after compression (\cref{sec:study-results-intuition-building}).
We demonstrate how \system can help practitioners identify and resolve breakages (\ref{challenge:internal-differences}) in the context of a generative language model for question answering.

We use an off-the-shelf \texttt{T5-Large} model~\cite{raffel2020exploring} that achieves an F1 score of 90.5\% on the Stanford Question Answering dataset~\cite{rajpurkar2016squad} (\cref{fig:squad-case-study}A).
The model's original performance is competitive with humans', but since the model is large (775 million parameters), we'd like to compress it to improve its speed and space utilization.
Following common compression workflows from our participants (\cref{sec:study-results}) and the literature (\cref{sec:related-work}), we apply magnitude pruning across all of the model's parameters.
However, this causes steep performance drops even at low levels of compression (e.g., 4\% F1 after pruning only 10\% of parameters).
Looking at the top changes in predicted answers in the \behaviors view (\cref{fig:squad-case-study}B), we see that magnitude pruning has broken the model's generation.
The 10\% pruned model repeats words from the context paragraph (\eg ``\texttt{Super Bowl LII LII LII ...}''), and the 30\% pruned model's output is meaningless (``\texttt{a a ...}'').


\system can help us understand why magnitude pruning has negatively affected the model.
There are many possible reasons this compression strategy could have failed\,---\,the model may have low compressibility, essential weights may have been inadvertently pruned, or magnitude pruning may not be well-suited to this task.
Since it is challenging to determine the cause using performance alone, we use the \layers view to inspect parts of the model that have been pruned.
Sorting the models' layers by how much their weights have changed, we see that the most changed layers are all normalization layers (\cref{fig:squad-case-study}C).
Normalization layers ensure a consistent activation distribution throughout the model, so over-pruning them can lead to unexpected behavior.
However, since the model has relatively few normalization weights, we would not necessarily expect magnitude pruning to have pruned them so aggressively.
To test if pruning the normalization layers caused the performance drop, we design a follow-up experiment that restores the normalization layers in the pruned models to match the original model, effectively unpruning them.
This leads to a full recovery in F1 for the 10\% and 30\% pruned models, indicating that pruning the normalization layers was a substantial issue in our original compression experiment.

We can also use \system to understand if the repaired models can be pruned any further.
To do so, we browse model activations in the \layers view for the original model, the 30\% pruned model with restored normalization layers (the fixed model), and the 50\% pruned model with restored normalization layers (a broken model). 
We observe that the outputs of the self-attention module have changed significantly during pruning, even in the working model, which may signify that the model is robust to changes in these modules. 
By pruning additional parameters from the attention modules, we can reduce the model size by roughly 30\% while achieving 83\% F1 score.
Further, reviewing these models' predictions relative to the base in the \behaviors view (\cref{fig:squad-case-study}D) confirms that their generative abilities are preserved.
Here, \system helped us debug our compressed model's low performance and design compression modifications that result in efficient, high-performing models.

\subsection{Discovering Compression Artifacts}
\label{sec:case-study-compression-artifacts}
Compression artifacts\,---\,changes in model behavior caused by compression\,---\,can subtly affect model quality, decrease edge case performance, and increase bias without reducing overall accuracy, making them difficult to detect yet crucial to address.
To demonstrate \system's ability to identify compression artifacts (\ref{challenge:behavior-differences}), we apply it to study known compression-induced biases in face classification models~\cite{liu2018large, hooker2020characterising}.
Following Hooker~\etal~\cite{hooker2020characterising}, we train a ResNet18~\cite{he2016deep} on CelebA~\cite{liu2018large} to predict whether each image has the attribute \texttt{blond}.
Then, we iteratively train and compress seven ResNet18~\cite{he2016deep} models on the same binary classification task using global magnitude pruning~\cite{zhu1710prune} at $10\%$, $30\%$, $50\%$, $70\%$, $90\%$, $95\%$, and $99\%$ final sparsity (see Supplementary Material Sec. \ref*{app:celeba-models}).
Each resulting model achieves similar accuracy on the test set, ranging from $87.4\%$ to $94.4\%$.

\begin{figure*}[t]
    \centering
    \includegraphics[width=\linewidth]{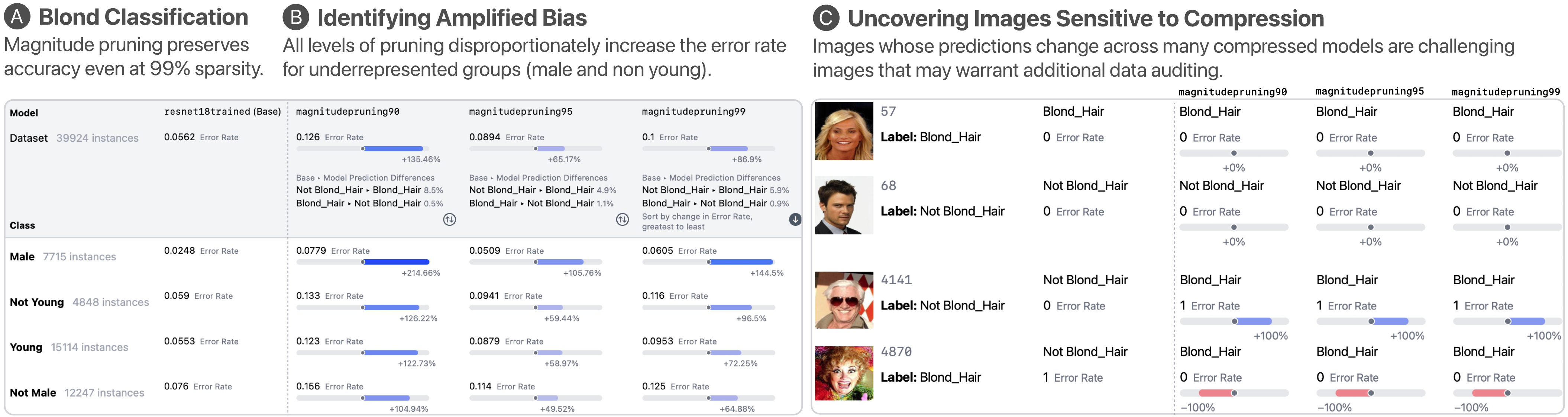}
    \caption{\system can help identify compression induced bias and perform data auditing. The \behaviors view reveals that compressing image classification models (A) disproportionately impacts rare classes (B) by forgetting hard-to-classify images (C).}
    \label{fig:celeba-case-study}
\end{figure*}

To identify potential sources of bias, we use \system to understand how the models' small drops in performance are distributed over the images.
In the CelebA dataset~\cite{liu2018large}, \texttt{male} and \texttt{not young} are underrepresented attributes.
If compression is introducing bias by forgetting rare classes, it will have a disparate impact on these images.
To inspect this, we compare the relative accuracy of the pruned models to the uncompressed model in the \behaviors tab (\cref{fig:celeba-case-study}B).
This view immediately surfaces that compression has disproportionately impacted the performance of rare classes.
While the 99\%-pruned model makes $64.9\%$ more errors for \texttt{not male} and $72.3\%$ more for \texttt{young}, in the rare classes, it makes $145.5\%$ more errors for \texttt{male} and $96.5\%$ for \texttt{not young}.
This is concerning since performance on underrepresented and difficult instances is a primary reason that ML practitioners start with a large model~\cite{liebenwein2021lost,li2020trainbig}. 
This may signal the need for further model development, compression experiments, and bias mitigation before we are comfortable using these models.
Previously, we would have needed to compute model performance on each attribute and interpret a table of relative percentages (such as in Hooker~\etal~\cite{hooker2020characterising}); however, using the visual affordances of \system, we are able to quickly identify bias without additional computation.

While \system supports bias identification using traditional error metrics, it can also audit bias in settings where we do not have access to data with sensitive attribute labels.
In this setting, we use \system to identify a subset of images that may merit additional inspection or human-in-the-loop annotation, similar to Compression Identified Exemplars~\cite{hooker2020characterising, hooker2019compressed}.
We begin by loading the compressed and uncompressed models into the \behaviors tab and sorting all images by the change in error rate between the uncompressed and $99\%$ compressed models (\cref{fig:celeba-case-study}C).
Many images are classified differently between the original model and the compressed models, indicating that these images are most sensitive to compression.
Looking closely at these images, we see they are often challenging images (\eg people with silver hair or hair occluded by headwear), confirming prior analysis~\cite{hooker2019compressed, hooker2020characterising} that images sensitive to compression are often challenging to classify.
In this way, \system supports data auditing to help users uncover instances sensitive to compression that may warrant further data cleaning and quality control.

\section{User Study with ML Practitioners}
\label{sec:evaluation}
To evaluate \system, we conducted a user study with \NumStudyParticipantsWord ML practitioners who work on model compression. Participants included research scientists developing new compression techniques, ML engineers compressing models for deployment, and software engineers building compression tools (\cref{tab:interview-participants}).
The goal of the study was to understand how participants would use \system to make sense of compression experiments, build intuition about compression strategies, and explore compressed model behavior.

\begin{table}
\centering
\caption{We evaluate \system with \NumStudyParticipants compression practitioners. While each participant uses compression in their work, they have diverse roles, ML applications, and compression workflows.}
\label{tab:interview-participants}
\resizebox{\columnwidth}{!}{%
\begin{tabular}{rlll}
\textbf{ID} & \textbf{Role} & \textbf{ML Application} & \textbf{Compression Workflow}\\
\midrule
    P1 & ML Engineer        & Efficient ML algorithms   & Compresses existing models \\
    P2 & ML Manager         & Efficient ML tooling      & Compresses existing models \\
    P3 & Research Scientist & Efficient ML tooling      & Compresses existing models \\
    P4 & ML Manager         & 3D computer vision        & Develops efficient models \\
    P5 & ML Manager         & 3D computer vision        & Develops efficient models \\
    P6 & Research Scientist & Efficient ML algorithms   & Develops efficient models \\
    P7 & Research Scientist & Multi-modal ML            & Develops efficient models \\
    P8 & Software Engineer  & Efficient ML tooling      & Builds compression tools\\

\end{tabular}
}
\end{table}

\subsection{Study Methods}
Each study session was conducted via video chat and lasted 45 minutes--1 hour.
We began by discussing the participant's current compression workflows and introducing them to \system.
Then, we asked participants to imagine they were part of a team compressing an existing image classification model and to think aloud as they used \system to complete two tasks:

\begin{enumerate}[label={\bfseries T\arabic*.}, ref={\bfseries T\arabic*},itemsep=1ex, labelindent=0pt, wide=0pt]

    \item First, we showed participants a ResNet50 image classification model~\cite{he2016deep,paszke2019pytorch,deng2009imagenet} and 19 of its compressed variants, including 8-bit quantization, global gradient and magnitude pruning at various sparsity levels, and post-pruning fine-tuning and calibration. We asked participants to identify successful experiments, explain why some experiments failed, and suggest future experiments. \label{user-study:task1}

    \item Next, we updated the interface to display all 52 models, including the 20 models from \ref{user-study:task1}, a MobileNet V2 model~\cite{sandler2018mobilenetv2} with global magnitude pruning, iteratively pruned and fine-tuned models, and models with combined compression techniques. We asked participants to evaluate the latency, size, accuracy, and behavior of these models and select the model(s) they felt confident deploying. \label{user-study:task2}
\end{enumerate}

To analyze the results of the study, two authors reviewed the video-recorded sessions.
They performed open coding to identify participants' insights during the tasks and their broader perspectives on compression and \system.
Using the resulting 357 codes, they conducted an iterative affinity diagramming and thematic analysis process to identify progressively higher-level themes.
This process generated 17 final themes that we use to structure \cref{sec:study-results}.

\subsection{Study Results}
\label{sec:study-results}

Overall, participants reported that \system provided a structured compression workflow, enabling them to perform analyses that would have been challenging and time-consuming with existing tools, build intuition about compression techniques that inspire new compression experiments, and identify subtle but problematic model behaviors introduced during compression.

\subsubsection{Unifying Disjoint Compression Workflows}
\label{sec:study-results-workflows}
Our user study participants reported needing multiple tools to execute their current compression tasks, leading to tedious back-and-forth analysis across tools that made it challenging to understand the overall impact of their compression experiments.
While each participant was knowledgeable about compression (average $3.5$ on a self-reported $1$--$5$ scale of compression experience) and used compression regularly (5/\NumStudyParticipants use compression in every project), there was not a standard set of tools used by all participants.
This inconsistency was often due to a lack of compression-specific tooling that supported the breadth of participants' tasks, including training and compressing models (P1--P8), evaluating their performance and behavior (P4, P6, P7), and ensuring they meet specified efficiency budgets (P1--P8).
As a result, participants often created custom tools, such as Jupyter Notebook charts for performance analyses (P1, P6, P8) and spreadsheets for model tracking (P3, P5), or they repurposed existing general model analysis tools to understand the performance of a compressed model (P1--P4).
When compression-specific tools existed, participants found them invaluable to their analysis (P3, P4).
However, these tools were often still specific to a particular aspect of compression analysis (\eg hardware performance~\cite{hohman2024talaria}).
As a result, participants found it challenging to perform comprehensive analysis and lamented that existing tools did not support many of their most critical tasks, like budgeting (P2), layer-wise analysis (P1), and compression-specific comparison (P2, P3, P4).

Unlike participants' existing workflows, with \system, critical compression tasks, like metric analysis and budgeting (\ref{challenge:budgeting}), experiment comparison (\ref{challenge:no-single-method}), and layer-wise behavioral inspection (\ref{challenge:behavior-differences} and \ref{challenge:internal-differences}) are all in one place:
\begin{quote}
    \textit{``[In] a lot of my typical workflows [...] you have to have 10 tabs in parallel in your browser and switch between them. I find that [\textsc{Compress and Compare}] is really bringing all the different aspects into a single view.''}~---~P2
\end{quote}
For instance, an approach used by 4/\NumStudyParticipants participants was to identify candidate models that fit their performance budget via the \filter, hone into the one or two best-performing models using the \scatterplot metrics, and search for patterns in the best models' compression recipes using the \modelmap.
Through this process, participants quickly identified ``deployable'' models that were small enough to fit on device while still maintaining task performance.
P1, P5, P6, P7, and P8's \overview analysis quickly converged from 52 models to a single 8-bit quantized ResNet50 model that reduced latency and size while nearly maintaining the uncompressed model's accuracy.
Having a comprehensive overview of compressed model variants gave participants confidence to pitch this model to their teammates and use it to design new experiments that could result in even greater efficiency.

The unified compression interface also sparked discussion about how \system could integrate into collaborative compression settings. 
Participants regularly collaborate with team members to complete their compression tasks, such as sending compressed models to QA specialists for targeted evaluation (P4, P5), negotiating resource budgets with product managers (P6, P8), and mentoring model developers on compression methods (P2).
However, it can be challenging to collaborate on a compressed model across various tools and with collaborators with varying skill sets.
In participants' current workflows, experimental results are distributed across tools, so participants were excited to use \system as a centralized communication tool.
P6 and P8 were interested in presenting their experiments to budget managers using \system to advocate for budget increases by interactively demonstrating how best-case model performance improves as the budget relaxes.
Participants also expressed interest in using \system to collaboratively compress models, such as by flagging potential compression-induced issues in the \behaviors view for review by their QA teammates (P4, P5) and setting up experiments in the \modelmap that demonstrate compression pitfalls to less experienced engineers (P2).
Whereas existing compression workflows tended to become ad hoc when experimenting with many different algorithms, techniques, and pipeline structures, participants were excited for \system to provide structure to the collaborative search for an efficient and accurate model.



\subsubsection{Building Intuition about Model Compression}
\label{sec:study-results-intuition-building}
Beyond simply selecting a desired compressed model, \system's visual and interactive components helped participants build intuition for how compression algorithms impact model performance and generate hypotheses about ways to improve future experiments.
Participants found the combination of the \modelmap and the \selection view to be an intuitive way to understand and reason about the space of compression experiments.
Viewing the columns in the \modelmap helped users understand the set of compression algorithms that had been applied (P3, P4, P7) and identify patterns in how the best-performing models were generated (P2, P7).
To dig into a particular pattern, participants would often run visual ``experiments'' by selecting a group of models within a region of the \modelmap (\eg all the magnitude pruned models) and comparing their metrics in the \selection view (P2, P4, P6, P8).
These in-depth explorations influenced participants' intuitions about how compression techniques affected their models more broadly.
For instance, by comparing the accuracy and efficiency of two quantized models, P4 identified that quantization preserved performance much better for a large ResNet50 (25.6M parameters) than for a smaller MobileNet V2 (3.5M parameters).
While P4 regularly uses quantization, this discrepancy in performance caused them to reflect that the success of quantization 
\textit{``is definitely dependent on the base model; if you want something to work for quantization, you have to start at the right place.''}
Finding the best compression technique is challenging
(\ref{challenge:no-single-method}), so building intuition for the types of models that benefit from quantization can help P4 design more effective compression recipes moving forward, such as those that only apply quantization to large models.

Building on their high-level understanding of the experimental space, participants used the \performancecomparison views to develop a deeper intuition about compression's impact on models' internal representations.
Using the \layers view, participants debugged subtle problems in compression experiments that led to poor model performance.
For example, P8 recognized that a particular model's \textit{``batch norms had been absolutely flattened''} by quantization.
Batch normalization layers can have a substantial impact on downstream performance because they set the output value ranges at each layer, so this finding led P8 to suggest \textit{``freezing all the batch norms''} during quantization as a way to maintain model performance.
By building intuition through their \layers exploration, participants ideated a range of subsequent experiments.
These next steps included combining current compression techniques in new ways (\eg combining magnitude and gradient pruning) and expanding the space of operations, such as by tailoring compression to specific layers of the network.
For instance, in their \layers analysis, P5 and P7 noticed earlier network layers had fewer parameters yet were pruned at the same rate as later layers.
They hypothesized that, since later layers have more parameters, they have more redundancy and could withstand greater compression rates, so they designed an experiment that pruned layers as a function of their number of parameters.
With \system, participants deepened their understanding of how model parameters respond to compression techniques (\ref{challenge:internal-differences}) and used their insights to generate new experiments that could lead to more efficient and accurate models.

\subsubsection{Encouraging Comprehensive Compression Analysis}
By interactively integrating behavioral analysis with traditional metric-based compression analysis, \system extended participants' existing behavioral analysis workflows and motivated them to consider the broader impacts of compression.
Evaluating compressed model behavior on held out data was a key aspect of some participants' workflows (P5--P8) because it helped them identify subtle but important behavioral changes (\ref{challenge:behavior-differences}):
\begin{quote}
    \textit{``When you compress a model you care about its quality on rare classes. The biggest risk when you compress your model is all of a sudden it becomes [problematic].''}~---~P8
\end{quote}
Participants used the \behaviors view to analyze model behavior across an entire dataset to ensure that compression had not induced biases or spurious correlations.
The ability to sort by relative change in correctness helped them identify classes that experienced the most errors and inspect individual instances that were misclassified.
This procedure revealed that many compressed models' mistakes were acceptable, such as mistakes on multi-object images (\eg \texttt{coffee pot} in a \texttt{stove} image) or related classes (e.g., \texttt{Great Dane} misclassified as another dog breed).
However, it also helped them uncover subtle patterns and identify potential compression-induced biases.
For instance, P5 sorted the \behaviors view by decreasing change in correctness and observed that \texttt{stove} images had lost 22\% accuracy, whereas overall the model only experienced a few percent decrease.
Inspecting the \texttt{stove} images whose classifications had changed revealed a spurious correlation between \texttt{stoves} and \texttt{microwaves}.
P5 worried that the compressed model could be relying on the presence of one to classify the other.
By viewing model behavior over an entire dataset, participants, like P5, were able to identify concerning patterns in the model's behavior, hypothesize reasons for the problem (\eg a disproportionate amount of training images contain both objects), and develop a plan to address them (\eg flagging these examples for QA team review).

While participants primarily analyze correctness in their current workflows, having access to additional plug-and-play metrics in the \behaviors view spurred new analysis processes.
During P1's behavioral analysis, they discovered that magnitude pruning resulted in a larger KL divergence in output probabilities than quantization.
While they do not use KL divergence in their standard analysis pipeline, having access to it in \system helped them distinguish between otherwise similar compressed models and select the one that best reflected the original model's outputs.
\begin{quote}
    ``\textit{I don't look at KL divergence very frequently, but KL divergence is zero for [the quantized] model and non-zero for [the pruned] model. There was only a 6\% accuracy regression [for the pruned model], so it's surprising. I'm impressed by the KL divergence [of the quantized model] being extremely low. It's the incumbent solution.}''~--~P1
\end{quote}
Similarly, P2 and P3 uncovered that magnitude pruning resulted in higher model confidence than quantization.
With the knowledge that these differences in model outputs existed, participants were able to generate hypotheses for their existence (\eg confidence increases may be a result of overfitting with fewer parameters (P2)) and strategies to account for them (\eg setting a different prediction threshold based on the compression algorithm (P3)).
Overall, \system extended participants' compression workflows to integrate behavioral analyses with their standard metric-based budgeting procedures. 
As a result, participants seamlessly switched between the two, iteratively selecting a candidate compressed model and interrogating its behavior to identify biases and hypothesize new compression experiments.

\section{Discussion}
\label{sec:discussion}
We present \system, an interactive visualization system for tracking and comparing compression experiments. 
Based on challenges experienced by real-world users, we design \system to support critical and unsupported compression analysis tasks, including managing interconnected compression experiments, interrogating the impact of compression on model behaviors, and ideating promising future compression experiments. 
Through case studies on generative language and image classification models, we demonstrate how our system helps users repair issues with compression and identify compression-induced bias.
Moreover, our user study with compression experts illustrates how \system shifts users' compression workflows from disjoint analysis across tools toward a single analysis platform that facilitates collaborative decision-making about model selection and exploration.
Here, we discuss the implications of our results for future ML development and evaluation tools, as well as the current limitations of our work and possible solutions.

\subsection{Designing Compression-Aware ML Workflows}
Throughout the design and analysis of \system, we encountered compression-specific challenges.
While compression analysis could be considered a special case of general ML evaluation, our user study participants struggled to extend traditional model evaluation workflows to their compression-specific tasks.
Our work suggests ways future tools can improve the overall ML development process by integrating compression considerations:

\vspace{1.2mm}
\textit{Bridging data- and model-centric evaluations.}
Most existing ML development tools either focus on data-centric evaluations of model accuracy and behavior~\cite{cabrera23zeno,boggust2022embedding,sivaraman2022emblaze,robertson2023angler} or architecture-specific evaluations of model internal layer characteristics~\cite{hohman2024talaria,olah2017feature}.
In contrast, we found that linking data-centric (\ie \behaviors) and architecture-specific visualizations (\ie \layers) helped users interpret the functional characteristics of model components, similar to systems like DeepCompare~\cite{murugesan2019deepcompare} and ActiVis~\cite{kahng2018activis}.
For instance, in our case studies and user studies, participants used the connection between the \layers and \behaviors views to identify that compressing batch normalization layers directly worsened the quality of the model's outputs.
By identifying a functional relationship between the model's architecture and its behavior, users were then able to ideate new, better-performing experiments (\eg removing compression on normalization layers).
By developing joint visualizations of evaluation data and model architectures, compression tools can help users connect aspects of a model's design to changes in its behavior.
    
\vspace{1mm}
\textit{Negotiating trade-offs between model quality metrics.}
Compression practitioners often trade off between model size, latency, and accuracy.
While existing ML pipelines have addressed similar issues between accuracy and fairness~\cite{valdivia2021fair} and accuracy across tasks~\cite{lin2019pareto}, interactive model selection tools have not explicitly explored helping practitioners make these trade-offs, \eg by identifying Pareto frontiers.
Further, our user study participants indicated that efficiency and accuracy budgets are often collaborative and malleable targets, as opposed to hard quantitative thresholds.
By visualizing experimental results and metric trade-offs, compression tools can help practitioners communicate their constraints and advocate for budget changes when needed.

\vspace{1mm}
\textit{Tracking model provenance during iterative development.}
Unlike model architecture and hyperparameter search, where experiments are often simple Cartesian products of several variables, compression experimentation is more readily modeled as a tree of ``recipes'' where nodes represent models and edges represent operations. 
Practitioners often start with a single model or a few related models that are known to perform well and apply varying compression recipes to them, creating the branching structure depicted in the \modelmap.
This process results in many interconnected models, and it can be challenging to keep track of the operations that created a particular model and its relationship to other models.
User study participants found visualizing models in this tree structure helped them build intuition for aspects of experiments that worked well and how future experiments may behave.
Future compression tools may consider a similar tree visualization or new ways to communicate model provenance to practitioners.

\vspace{1mm}
\textit{Comparing complex differences across many models.}
Existing tools (including those for compression) tend to focus on profiling and evaluating a single model~\cite{hohman2024talaria,cabrera23zeno,bauerle2022symphony,schlegel2022vinnpruner}. 
In contrast, our study underscores the value of directly supporting comparison~\cite{gleicher2017considerations,sivaraman2022emblaze}.
By visually juxtaposing metrics, predictions, and internals from several models, \system reduces the cognitive load required to determine which differences are most actionable. 
Our system enables workflows that rapidly transition between \textit{comparative relationships}, ranging from black-box comparisons of top-level metrics to comparisons of individual layer activations. 
Future tools can prioritize comparison and look to practitioners' existing comparison strategies to understand when linking multiple comparative relationships leads to productive insights.

\vspace{1.2mm}
These compression-specific challenges provide a basis for the design of future compression-vis tools, but they could also suggest ways to improve general ML analysis workflows.
For instance, our user study participants found the \modelmap tree structure to be highly intuitive, and some suggested applying it to other stages of model development (P4, P18, P15), such as creating a timeline-based \modelmap that organized model results based on when a user ran each experiment.
Moreover, extending the layer-wise activation comparison in the \layers view could support complex hyperparameter search workflows.
Additionally, features demonstrated in \system may be worth integrating into general-purpose ML tools, helping those tools cover a broader range of existing workflows and encouraging practitioners to focus on efficiency earlier in the development process.


\subsection{Limitations and Future Work}
We designed \system to support practitioners current compression workflows; however, as model compression becomes a more established and standardized discipline, it is possible that many of the iterative workflows we observed in this study will be superseded by automated approaches. 
However, interactive visualization has potential benefits for compression work even if automated approaches eventually become standard. 
First, even though effective AutoML strategies exist, these techniques often still require data scientists to invest considerable time to distill model behavior into a single objective function~\cite{karmaker2021automl}. 
Second, \system seeks not only to help practitioners produce the best compressed model, but also to help them build intuition about compression techniques and how they affect model characteristics. 
Just as prior work visualizing ML model internals and behaviors~\cite{karpathy2015visualizing,reif2019visualizing,zeiler2014conv} has allowed people to understand how these models work, a better collective understanding of compression can make these techniques more accessible and inspire new approaches.


Our user study and prior formative research primarily interviewed compression experts and ML engineers who were already well-versed in model compression. 
While this allowed us to get the most relevant feedback on how to design compression-specific systems, our system and takeaways do not consider the needs of novice users.
It is likely \system could be extended to support ML practitioners less accustomed to compression, such as by providing suggestions on what techniques to try or incorporating a graphical interface to run compression experiments.
Such adaptations would empower non-experts to apply compression, but they could also support compression experts in running on-the-fly experiments based on their insights from \system, such as removing compression from a specific layer or making a slight change to the sparsity value.


Further, several participants suggested extensions to \system that could help it better match the specific needs of their teams, including supporting very large datasets and models as well as displaying custom efficiency metrics. 
Although the tool currently supports running model computations remotely or in advance, it does require the models being compared to be loaded simultaneously in memory so that each layer can be compared one-at-a-time. 
More efficient comparison techniques that do not require jointly loading several models may help \system scale more easily.

Finally, a key requirement of \system is its integration between the visualization system's and a user's model development code. 
However, like many other prototype tools for ML development, the capabilities that make \system a versatile tool can also necessitate considerable set-up work, particularly for custom model architectures. 
User study participants noted that the potential difficulty of importing models into the tool could be a critical hurdle to accepting it into their workflow. 
Future work is needed to design code-level interfaces that link \system into the tools practitioners already use to develop models.

\section{Conclusion}
\label{sec:conclusion}

Making ML models smaller, faster, and more energy-efficient can enable exciting new user experiences and broaden access to existing ones. 
Our work forms part of a nascent body of literature on facilitating the process of compressing models, which can help make these use cases practical as models become increasingly large and powerful. 
Through the design and evaluation of \system, we aimed to understand and address challenges in ML model development made salient by compression, including the need for extensive iteration, human-centered trade-offs between user experience metrics, and subtle changes in model behavior. 
Future compression-focused data visualization research can continue to make creating efficient ML models easier for a wider range of developers and teams, helping them make the experiences they envision a reality.

\acknowledgments{%
	
We thank our colleagues at Apple for their guidance on this research and our case study participants for generously sharing their time and knowledge with us.
We especially thank Guillaume Seguin and Xiaoyi Zhang for lending their expertise on compression and encouraging us to pursue this line of research.




}

\bibliographystyle{abbrv-doi-hyperref}

\bibliography{24-comparison-vis}



\end{document}